\documentclass[fleqn,10pt]{wlscirep}
\usepackage[utf8]{inputenc}
\usepackage[T1]{fontenc}
\usepackage{graphicx}
\usepackage{siunitx}
\usepackage{tabularx}
\usepackage{amsmath}
\usepackage{amssymb}
\usepackage{bm}

\DeclareSIUnit\gauss{G}
\usepackage{csquotes}

\title{Leveraging Prior Mean Models for Faster Bayesian Optimization of Particle Accelerators}

\author[1,*]{Tobias Boltz}
\author[2,*]{Jose L. Martinez}
\author[3]{Connie Xu}
\author[4]{Kathryn R. L. Baker}
\author[1]{Zihan Zhu}
\author[1]{Jenny Morgan}
\author[1]{Ryan Roussel}
\author[1]{Daniel Ratner}
\author[2]{Brahim Mustapha}
\author[1]{Auralee L. Edelen}

\affil[1]{SLAC National Laboratory, Menlo Park, 94025, USA}
\affil[2]{Argonne National Laboratory, Lemont, 60439, USA}
\affil[3]{Duke University, Durham, 27708, USA}
\affil[4]{ISIS Neutron and Muon Source, STFC, Oxfordshire, OX11 0QX, UK}

\affil[*]{tboltz@slac.stanford.edu, jl.mrtnz.mrn@gmail.com}

\begin{abstract}
Tuning particle accelerators is a challenging and time-consuming task that can be automated and carried out efficiently using suitable optimization algorithms, such as model-based Bayesian optimization techniques. One of the major advantages of Bayesian algorithms is the ability to incorporate prior information about beam physics and historical behavior into the model used to make control decisions. In this work, we examine incorporating prior accelerator physics information into Bayesian optimization algorithms by utilizing fast executing, neural network models trained on simulated or historical datasets as prior mean functions in Gaussian process models. We show that in ideal cases, this technique substantially increases convergence speed to optimal solutions in high-dimensional tuning parameter spaces. Additionally, we demonstrate that even in non-ideal cases, where prior models of beam dynamics do not exactly match experimental conditions, the use of this technique can still enhance convergence speed. Finally, we demonstrate how these methods can be used to improve optimization in practical applications, such as transferring information gained from beam dynamics simulations to online control of the LCLS injector, and transferring knowledge gained from experimental measurements across different operating modes, such as accelerating different ion species at the ATLAS heavy ion accelerator.
\end{abstract}

\begin{document}

\flushbottom
\maketitle
\thispagestyle{empty}

\section*{Introduction}
Particle accelerators are complex machines that have hundreds of free parameters which can be tuned to increase accelerator performance and beam quality. Controlling these parameters in real time during accelerator operations through the use of advanced algorithms, so-called online control, enables more complex beam manipulation capabilities and can help reduce the time spent on conducting routine tuning tasks. However, using optimization algorithms to solve facility-scale optimization problems that contain hundreds of free parameters is an open challenge due to the exponential scaling of parameter space volume with increasing numbers of parameters, often referred to as the \enquote{curse of dimensionality}~\cite{bellman_dynamic_1966}. When developing effective algorithms for accelerator control, it is important to incorporate prior information about beam physics into the algorithm to efficiently find solutions in extremely large parameter spaces.
Additionally, many accelerator facilities frequently switch beam conditions for different users. This is the case, for example, at both the Linac Coherent Light Source (LCLS) and the Argonne Tandem Linear Accelerator System (ATLAS), a heavy ion accelerator. Injector tuning at LCLS by hand can often take hours, both for restoring the beam after a shut-down or when switching between different beam conditions (e.g. adjusting the injector for different beam charge setups). While these configurations can be treated as separate optimization problems, the time needed to switch between conditions could be reduced by leveraging information from related machine configurations.

Bayesian optimization (BO) is one such algorithm type that has been used in accelerators and can incorporate prior physics information into the decision-making process~\cite{roussel_bayesian_2024}. BO methods have been shown to reduce the tuning time in several applications when compared to more traditional methods (e.g. see ~\cite{roussel_bayesian_2024,hanuka_physics_2021,duris_bayesian_2020}). The BO algorithm builds a Gaussian process model (GP)~\cite{rasmussen_gaussian_2006} that uses Bayesian statistics to predict objective function values (and corresponding uncertainties) from measurements, while factoring in a prior belief about the objective function. This model is then passed to an acquisition function to decide the next point to be observed in search of a globally optimal parameter set. Popular acquisition functions, such as Expected Improvement (EI) and Upper Confidence Bound (UCB)~\cite{brochu_tutorial_2010}, strike a balance between choosing points that add more information to the global GP model (exploration) and choosing points that are near predicted extremum (exploitation).

\begin{figure*}[t]
    \centering
    \includegraphics[width=\linewidth]{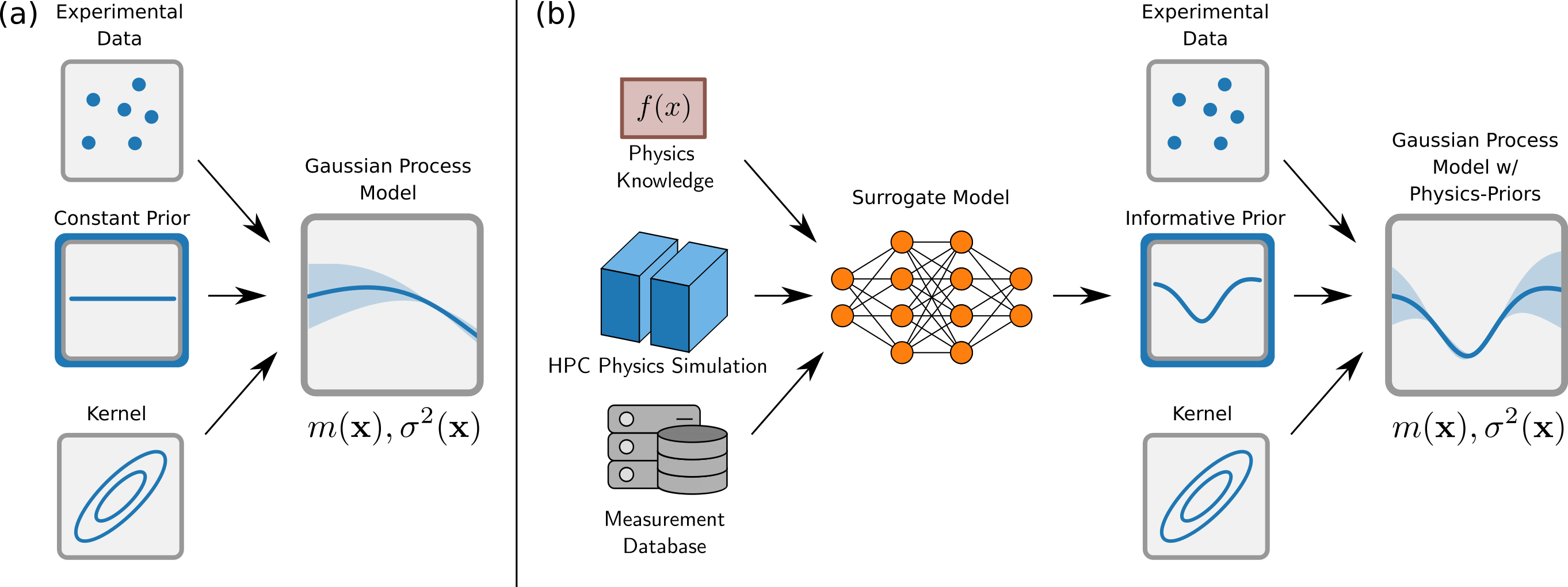}
    \caption{Illustration of how surrogate models can be incorporated into Gaussian process models used in Bayesian optimization algorithms for online accelerator optimization. (a) Conventional modeling approaches combine measurement data with an uninformative constant prior function and a kernel function to make predictions. (b) Fast-executing surrogate models trained using physics principles, physics simulations on high performance computing clusters, and historical data can be used in GP models as an informative prior function, which in turn improves the accuracy of model predictions, resulting in faster convergence of BO algorithms to optimal operating conditions.
}
    \label{fig:prior_mean_cartoon}
\end{figure*}

Previous work~\cite{duris_bayesian_2020,hanuka_physics_2021,bonilla_multi-task_2007} has demonstrated that including prior information from physics into BO algorithms increases their ability to efficiently converge to optimal solutions in high-dimensional parameter spaces. The use of Bayesian statistics (which inherently takes prior information into account) to build the GP model provides a natural framework for incorporating physics information into the modeling of accelerator-related optimization objectives. For example, GP models are built using a prior mean function, which specifies the prediction of the GP model in the absence of data. These prior mean functions are usually specified to be uninformative, as shown in Fig.~\ref{fig:prior_mean_cartoon}(a). However, if prior information about the objective function is known, either from beam dynamics laws, simulations or prior data, an informative prior mean can be specified to incorporate this information directly into the GP model, as shown in Fig.~\ref{fig:prior_mean_cartoon}(b). This can increase the accuracy of the GP model which helps guide the search for an optimum, improving the algorithm's ability to converge to solutions quickly in high-dimensional spaces.

In this work, we demonstrate and evaluate the use of fast-executing neural network (NN) surrogate models trained on beam dynamics simulations and historical data sets, as informative, physics-based prior mean functions in GP models to improve online accelerator control for two impactful use cases: injector tuning for the LCLS and tuning for different ion species at ATLAS. In particular, we demonstrate applications of this method in the context of realistic accelerator operations, including adapting surrogate models trained on beam dynamics simulations to online control at the LCLS injector, and transfer learning between different ion species at the ATLAS heavy ion accelerator, with improvements in convergence speed. These two cases were carefully chosen based on the expected benefit of the NN prior approach. The LCLS has to frequently switch setups for different users (different beam energies, bunch lengths, etc), and the injector is one of the more challenging parts of the machine to optimize due to space charge forces and the resulting sensitivity of the system to changes in controlled and uncontrolled inputs (such as drift in the initial laser spot). The injector is also not continuously tuned, because the relevant diagnostics are destructive. This makes it appealing both for using BO and for leveraging NN priors obtained from simulations and historical data. In addition, detailed physics simulations and ML-based surrogates of the LCLS injector are readily available. Similarly, ATLAS has to switch frequently between different ion species. Complicating this, beam dynamics for heavy ion linacs are very complicated and sensitive to machine inputs and changing conditions. Leveraging information across different configurations to speed up switching between ion species would substantially aid operation. The work presented here expands on previous work~\cite{xu_neural_2022,hwang_prior_2022,martinez-marin_real_2023} that was conducted in a more limited scope. The first shows an offline proof-of-concept  demonstration of using a NN prior for emittance and matching tuning on the LCLS injector. The second is a preliminary experimental demonstration at FRIB using an NN prior from historical data for tuning. The third is an initial experimental demonstration at ATLAS.
We also explore the impact that the accuracy of the prior model has on the speed of convergence, and present a method to counteract the effects of an inaccurate model. As physics models often deviate substantially from real machine behavior, handling of machine-model mismatch is a major consideration for practical use of the NN prior approach. 
We show that combining approximate or well-calibrated NN models with GP models is an effective, computationally inexpensive way to incorporate detailed physics information into BO algorithms, significantly reducing the number of optimization steps needed to converge to a solution, even in cases where the NN model does not perfectly predict the experimental objective. We also examine the impact of different levels of mismatch between the NN model and the accelerator on optimization speed, which is an important consideration. We find a high degree of model accuracy is not required; even coarse models still provide a benefit to BO convergence speed.
Our demonstration also shows the utility of using prior information across ion species to improve tuning speed, which is important for current and future nuclear physics machines.

\section*{Results}
Here we present simulated and experimental studies, comparing the optimization performance of BO algorithms with and without using NN models as prior means in GP models.

\subsection*{Simulation Studies}
\begin{figure}
\centering
\includegraphics[width=0.75\linewidth]{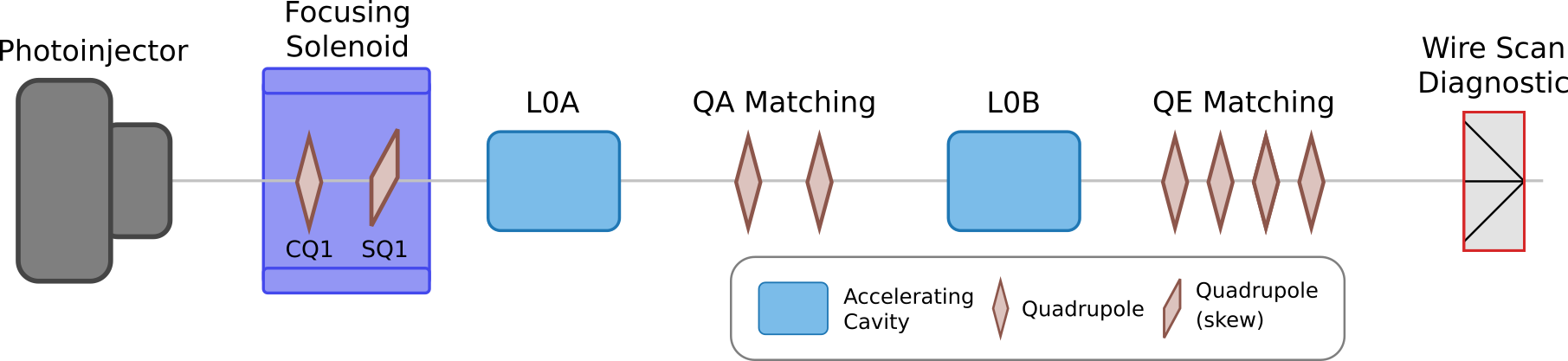}
\caption{Layout of the LCLS photoinjector beamline, including the photoinjector, magnetic focusing and matching elements, accelerating cavities, and a wire scan diagnostic to measure the beam profile in the transverse directions ($x,y$).}
\label{fig:lcls_injector_beamline}
\end{figure}

In order to understand the influence of a non-constant prior mean on BO performance, its effects are initially studied in simulations using a surrogate model for the LCLS photoinjector, the layout of which is shown in Fig.~\ref{fig:lcls_injector_beamline}. The beamline is parameterized by 16 scalar parameters that control the laser spot size on the cathode, the laser pulse length, the bunch charge, the phases and amplitudes of two RF accelerating cavities, and the focusing strengths of a number of solenoid, normal quadrupole and skew quadrupole magnets. We use a fully connected NN surrogate model trained on data from the beam dynamics simulation code IMPACT-T~\cite{qiang_impactt_2006} to predict five scalar quantities of the beam distribution (beam sizes and transverse emittances) at an optical transition radiation diagnostic screen downstream of the photoinjector (OTR2). This surrogate model is incorporated into the GP model built by Xopt~\cite{roussel_xopt_2023} as a prior mean function using the LUME-Model~\cite{mayes_lightsource_2021} framework to perform calibration and scaling to experimental units.

We run BO to minimize the transverse beam size of a round beam with respect to $\num{13}$ of the above machine parameters that control the magnet and RF settings (with the laser spot size, laser pulse length, and bunch charge parameters held fixed) using the objective function
\begin{equation}
    f_\text{LCLS}(\mathbf{x}) = \sqrt{\sigma_{x}^2 + \sigma_{y}^2} + | \sigma_x - \sigma_y | ~,
    \label{eq:lcls_objective}
\end{equation}
where $\sigma_x$ and $\sigma_y$ denote the beam size in their respective dimension. While the original model is used to represent the objective function $f_\text{LCLS}(\mathbf{x})$, the prior mean function $m(\mathbf{x})$ is given by variations of the same model with different levels of accuracy relative to the original.
This is a more realistic scenario than the particular case of a perfect prior mean model, i.e., a model $m(\mathbf{x})$ that perfectly predicts the objective function $f(\mathbf{x})$. In the latter case, the optimization problem becomes trivial and is typically solved within a single step.
The model accuracy is reduced by training a randomly initialized neural network with the same architecture as the LCLS surrogate on a grid of $3^{16} \approx \num{4.3e7}$ data points while limiting the number of training epochs. Models that are trained over fewer epochs show significant disagreement with respect to the LCLS surrogate, while models trained over more iterations tend to be more accurate. We generally use the Pearson correlation coefficient $r$ and the mean absolute error (MAE), which are detailed in the Methods section, to describe the model accuracy.

\begin{figure}
\centering
\includegraphics{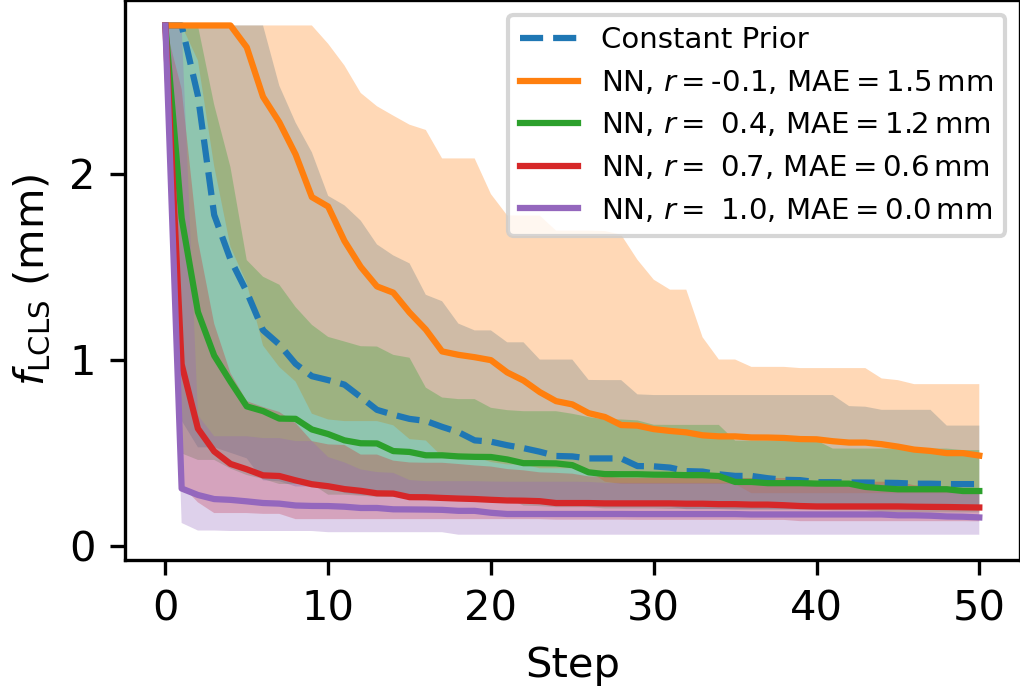}
\caption[LCLS BO with Imperfect Models]
{Beam size optimization using the LCLS injector surrogate model with different prior mean functions. Solid and dashed lines depict the median of the best seen value and the shaded areas the corresponding $\SI{90}{\percent}$ confidence level across $\num{100}$ runs.}
\label{fig:LCLSBOwithImperfectModels}
\end{figure}

The simulation results obtained by using these different models as prior mean functions for BO with the Expected Improvement (EI) acquisition function are summarized in Fig.~\ref{fig:LCLSBOwithImperfectModels}. Compared to a constant prior mean, NN prior models that have a strong positive correlation coefficient significantly improve the performance of BO during the first ten steps of optimization and eventually find a better optimum than generic GP models, despite having a MAE that is larger than the observed function value. On the other hand, the prior mean model with a negative correlation performed significantly worse than the constant prior mean. Overall, these results demonstrate how incorporating even imperfect NN models as GP priors can yield significantly better BO performance compared to those using a constant prior mean.

\subsubsection*{GP Prediction Scaling with NN Priors}
\begin{figure}
    \centering
    \includegraphics{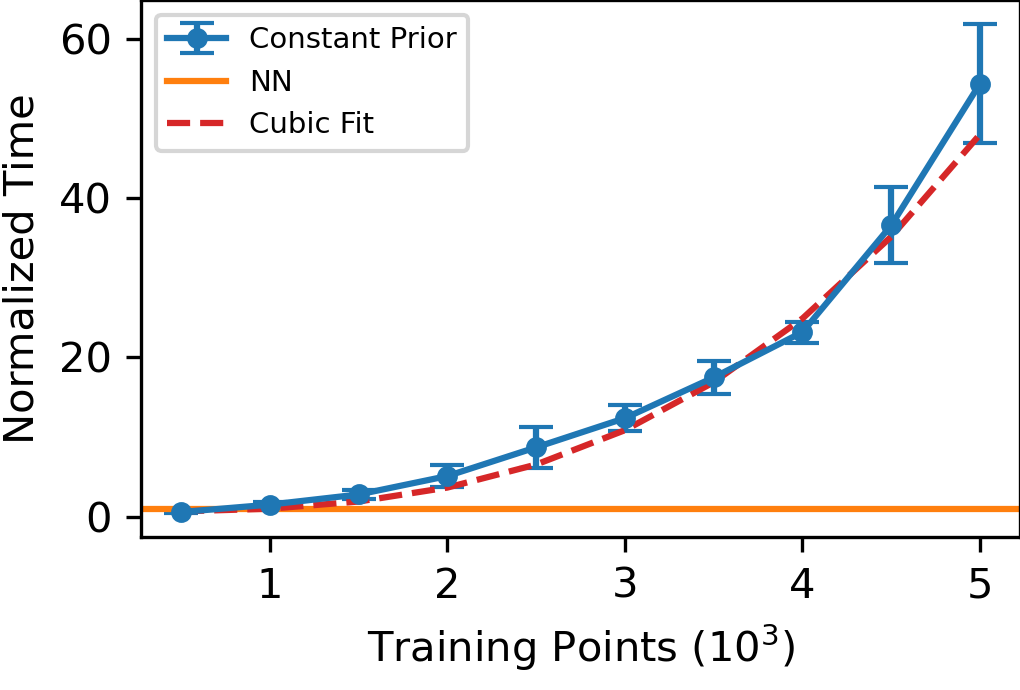}
    \caption{Scaling of computation time for GP evaluations on $\num{100}$ test points. While the computational cost for standard BO increases rapidly with the number of training points, inference time for the NN prior model is constant. The time axis is normalized by the NN inference time. These results were produced on an Apple M1 Pro @$\SI{3.20}{\giga\hertz}$ with $\num{8}$ physical cores.}
    \label{fig:evaluation_times}
\end{figure}

A critical advantage of NN models as prior mean functions is their relatively short evaluation time, which is independent of the training data set size. This enables incorporating extremely large amounts of data into the model from historical measurements or simulations, thereby circumventing the poor computational scaling of GP models. This is demonstrated in Fig.~\ref{fig:evaluation_times}, which compares the computational cost of evaluating GP models on a fixed number of points with different methods of incorporating prior data into the GP model. If we were to directly incorporate large training data sets into the GP model, the evaluation time scales as $\mathcal{O}(n^3)$. However, if we use a neural network to represent the prior information in the data set the evaluation time of the GP model is substantially smaller and independent of the number of training points. This computational scaling makes a significant practical difference when using these algorithms for online optimization, especially at user facilities that are in high-demand and need to spend as little time as possible tuning the beam.

\subsubsection*{Scaling up to High-Dimensional Problems}
\begin{figure}
    \centering
    \includegraphics{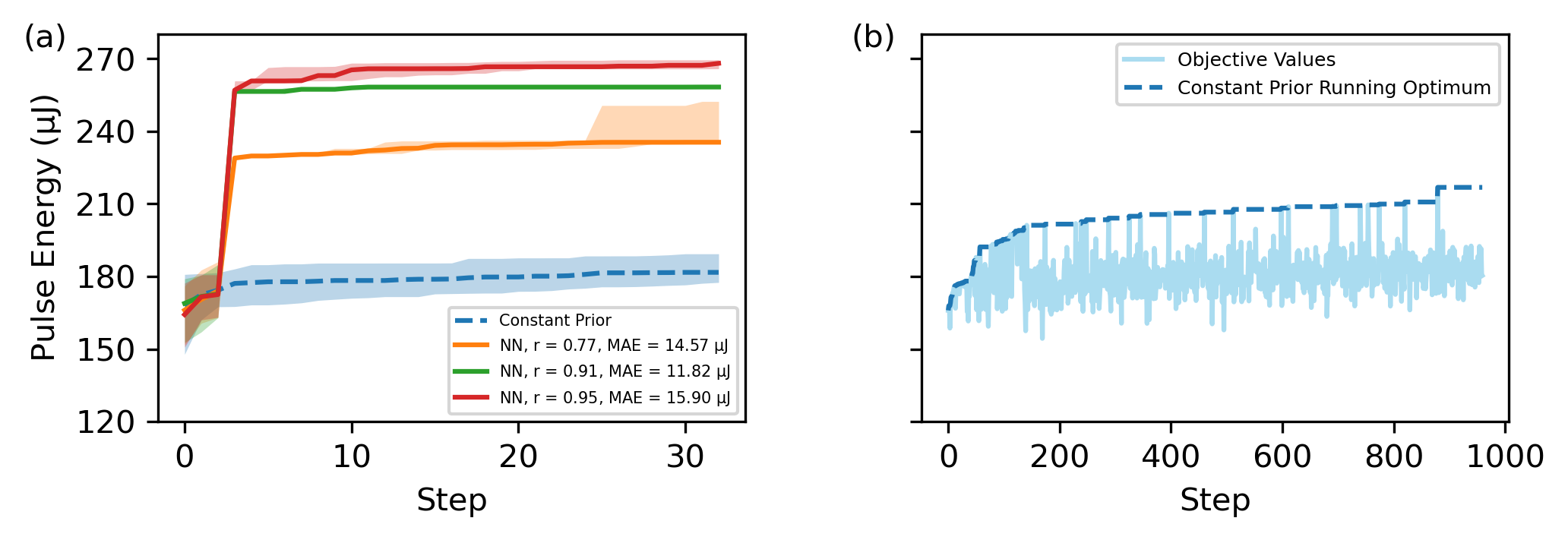}
    \caption{ (a) FEL pulse energy optimization with different prior mean functions. Solid and dashed lines depict the median of the best seen value and the shaded areas the corresponding $\SI{90}{\percent}$ confidence level across $\num{30}$ runs. The first $\num{3}$ steps correspond to random samples used to initialize the respective version of BO. (b) Illustration of an extended run of standard BO on the same problem. Even after almost $\num{1000}$ steps, the best seen value is still far from the optimum.}
    \label{fig:HighDimensionalBO}
\end{figure}

The performance benefits of using a NN prior mean model, as observed in Fig.~\ref{fig:LCLSBOwithImperfectModels}, stem from the fact that an informative prior helps to guide the algorithm to the most relevant areas of the parameter space without as much initial exploration. This additional guidance of the BO sampling sequence is especially interesting, and most impactful, when the optimization is performed over very large parameter spaces where an uninformed naive search can take prohibitively long to reach satisfactory results. This effect is demonstrated in a larger simulation of the LCLS beamline where the machine is parametrized by a total of $\num{73}$ quadrupoles and the objective is the Free-Electron Laser (FEL) pulse energy. The NN prior mean model is trained on data from the beam optics tracking simulations based on Cheetah~\cite{kaiser_cheetah_2024} and a FEL surrogate model developed using the simulation results from GENESIS~\cite{reiche_genesis_1999}. We then run BO with different models as prior mean functions to optimize over $\num{73}$ input parameters while evaluating against the beam physics model based on Cheetah and the FEL surrogate. The results shown in Fig.~\ref{fig:HighDimensionalBO} again demonstrate the improved performance that can be obtained by using an informative prior mean model, especially a high-correlation model that can facilitate more effective navigation of the high-dimensional parameter space. In contrast, standard BO takes significantly longer to converge to less satisfactory values while the NN prior based approach gets close to the optimum after only a few steps. To see how long it would take regular BO to converge, we ran it on the same problem. Even after $\num{900}$ steps, it has not yet reached the same performance the prior mean BO had after one initial step.

\subsubsection*{Performance Recovery for Low-Quality Prior Mean Models}
As it can be challenging in practice to obtain models with high or even medium accuracy, we also explore a few simple adjustments to quickly recover BO performance in cases where the model is found to be inaccurate. One of the main advantages of using the NN prior model seems to be that it can lead to better starts, that is, good albeit still sub-optimal values are typically found within fewer optimization steps. Convergence on the other hand seems to suffer much quicker under the biased search stemming from an inaccurate prior mean model. One straight-forward adjustment is thus to restrict the use of the NN model to a few initial steps, and to subsequently revert to the standard constant prior mean. For a more gradual transition we introduce a linear weighting factor
\begin{equation}
    m^{\prime}(\mathbf{x}) = w \, m(\mathbf{x}) + (1 - w) \, \text{const.}  ~,
    \label{eq:flatten}
\end{equation}
with $w \in [0, 1]$ and $w \rightarrow 0$ as the number of steps increases. Intuitively, one might think about this as a continuous \enquote{flattening} of the prior mean function as more steps are taken until it eventually becomes fully constant.

\begin{figure}
\centering
\includegraphics{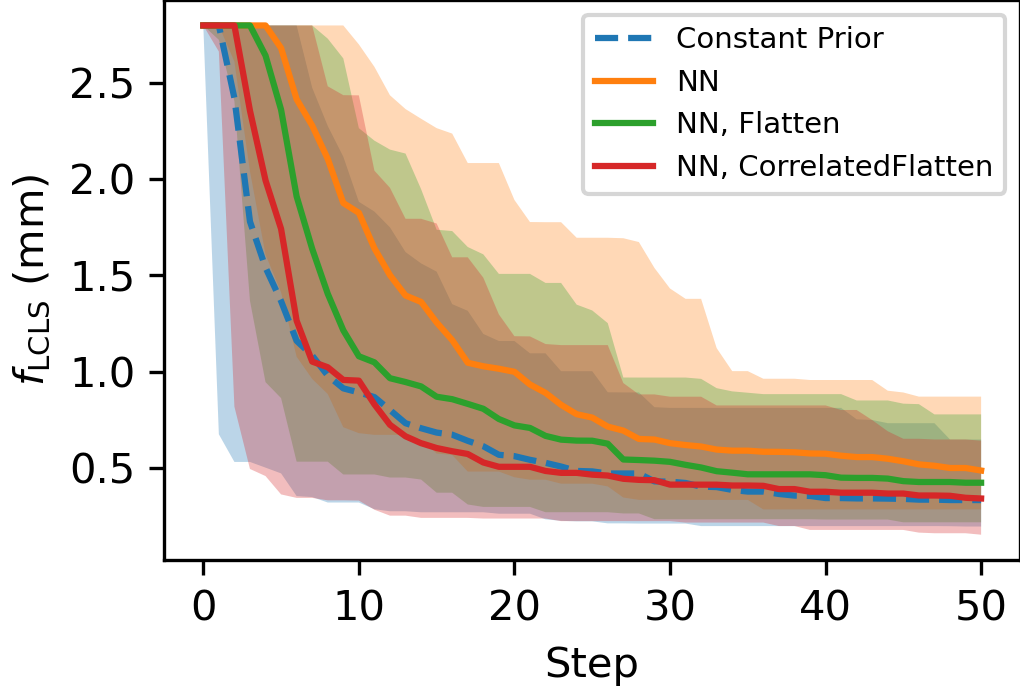}
\caption[LCLS BO Recovery]
{Recovery of BO performance for NN prior model with $r=\num{-0.13}$ and $\text{MAE}=\SI{1.50}{\milli\meter}$. Solid lines depict the median of the best seen value and the shaded areas the corresponding $\SI{90}{\percent}$ confidence level across $\num{100}$ runs.}
\label{fig:LCLSBORecovery}
\end{figure}

Based on the general shape of the performance curves in Fig.~\ref{fig:LCLSBOwithImperfectModels}, we linearly decrease the weight from an initial $w=\num{1}$ to $w=\num{0}$ at step $\num{10}$. As shown in Fig.~\ref{fig:LCLSBORecovery} (green line), this can improve the robustness of Bayesian optimization with a non-constant prior mean, bringing performance closer to that with a constant prior. To take this idea one step further, we let the weighting factor be chosen based on the Pearson correlation coefficient $r$ determined across previous samples (see Methods section)
\begin{equation}
    w = \text{clip}(r - w_0, 0, 1) ~,
    \label{eq:correlated_flatten}
\end{equation}
with the fixed offset $w_0 > 0$ serving as a threshold to disregard particularly bad models and a free parameter to further balance the trade-off between the NN model and a constant prior mean. With $w_0 = \num{0.2}$, this seems to further improve robustness and brings performance relatively close to that with a constant prior mean (red line) even if the NN model is very inaccurate. On the other hand, scaling the weighting factor with the observed correlation coefficient ensures good models are still identified and retains improved BO performance in those cases.

\subsection*{LCLS Experimental Demonstration}
\begin{figure}
\centering
\includegraphics{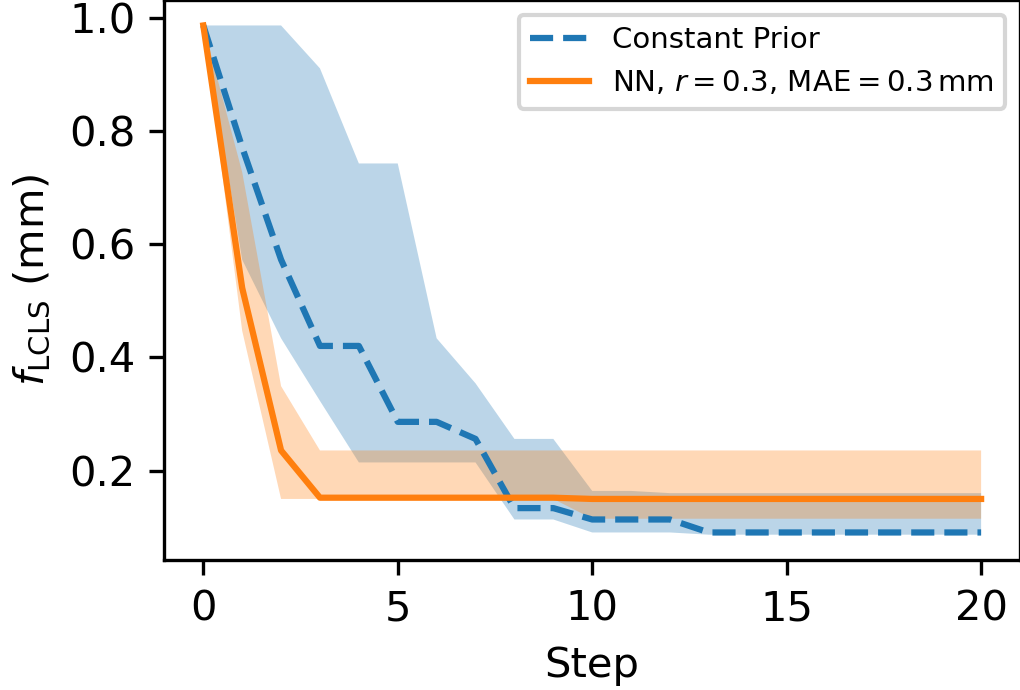}
\caption[LCLS BO with NN Prior]
{Beam size optimization at the LCLS injector with different prior mean functions. Solid and dashed lines depict the median and the shaded areas the $\SI{100}{\percent}$ confidence level across $\num{4}$ runs.}
\label{fig:ExpLCLSBOwithNNPrior}
\end{figure}

The LCLS surrogate model trained on data generated by the LCLS IMPACT-T model was used as an NN prior mean to perform online optimization of the LCLS photoinjector beamline. We repeatedly performed BO with the EI acquisition function starting with a fixed set of three randomly chosen initial points, with and without the NN prior mean function, to minimize the objective function in Eq.~\ref{eq:lcls_objective}. During this, the laser spot size, pulse length, RF cavity parameters and bunch charge were fixed, resulting in a 9-dimensional optimization problem. Results from these optimization runs can be seen in Fig.~\ref{fig:ExpLCLSBOwithNNPrior}. We observe that using the NN prior model consistently outperforms the constant prior in the first optimization steps. It takes the constant prior version five steps longer on average than the NN prior version to reach the same optimum. However, after approximately ten steps, the constant prior model surpasses the optimum achieved by the NN prior model, likely due to its low correlation with the measured objective, $r=\num{0.29}\pm\num{0.06}$, $\text{MAE}=\num{0.25}\pm\num{0.01}$, where $r$ denotes the Pearson correlation coefficient and $\text{MAE}$ the mean absolute error as detailed in the Methods section. The NN prior mean model provides an advantage in speed during the early stages of coarse optimization, and it is then surpassed in later refinement of the optimum by standard BO. Based on the simulation results discussed above, it is expected that NN models that are better calibrated to experimental data will increase the performance of BO in this context. In the meantime, it is possible to improve the optimization performance at later steps by transitioning from the NN model to a constant prior mean during the optimization, as demonstrated in the simulation section.

\subsection*{ATLAS Experimental Demonstration}
The Argonne Tandem Linear Accelerator System is a US Department of Energy User Facility dedicated to the study of low-energy nuclear physics with heavy ions. This facility operates year-round, using three ion sources, and serving at least six different target stations. The ATLAS linac undergoes reconfiguration or re-tuning once or twice a week, over $\num{40}$ weeks annually, delivering different ion beams and facilitating experiments in the different target areas.  The beam energies range typically from $\num{1}$ to $\SI{15}{\mega\electronvolt}/\text{nucleon}$. In recent years, the application of Bayesian optimization with Gaussian processes has gained traction at ATLAS, promising significant reduction in the time required to tune the accelerator~\cite{mustapha_machine_2022}.

\begin{figure}
\centering
\includegraphics[width=0.5\linewidth]{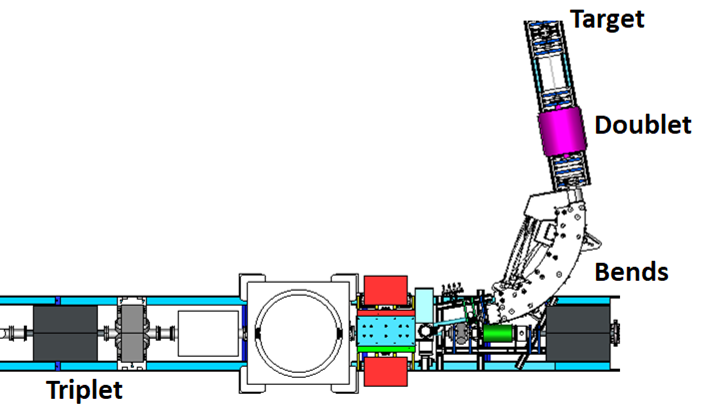}
\caption{Schematic of the AMIS beamline at ATLAS with controlled elements labeled.}
\label{fig:amis_beamline}
\end{figure}

\begin{figure}
\centering
\includegraphics{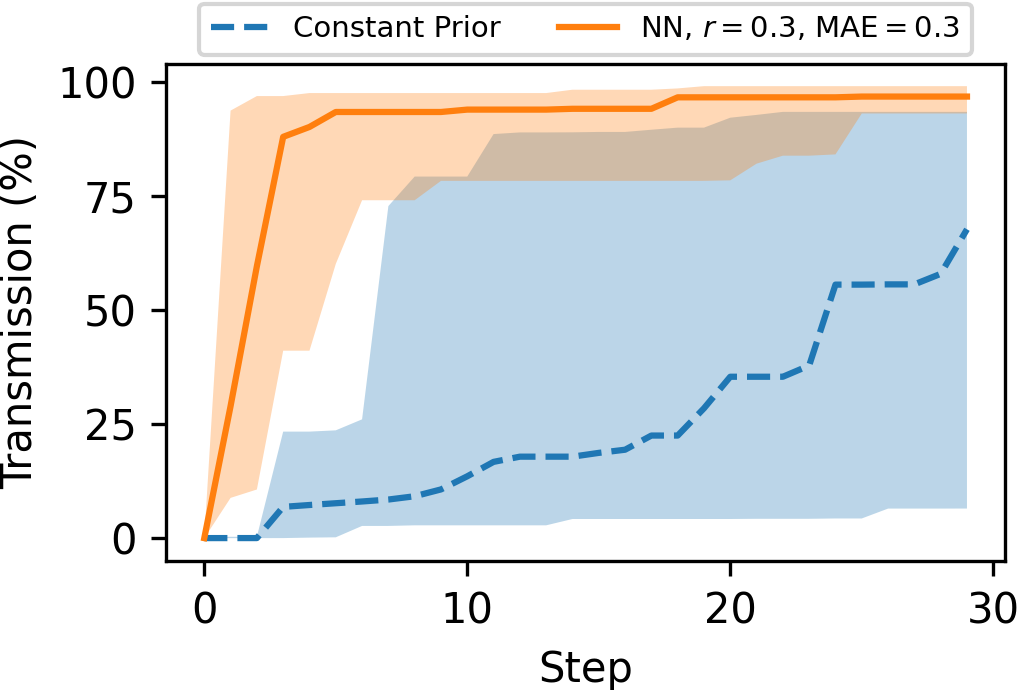}
\caption[ATLASnn-prior]
{NN prior versus standard BO with no prior knowledge. Solid and dashed lines denote the median best observed transmission and shading denotes the $\SI{90}{\percent}$ confidence level over $\num{20}$ runs.}
\label{fig:ATLASnn-prior}
\end{figure}

The main objective of beam optimizations at ATLAS is to maximize beam transmission to the target while preserving overall beam quality. In this application, the focus is to maximize beam transmission through the ATLAS material irradiation station (AMIS) line, shown in Fig.~\ref{fig:amis_beamline} by adjusting the current settings of a magnetic quadrupole triplet and doublet, totaling five parameters. We began with establishing the baseline performance of standard BO with a constant prior mean, initiating the optimization with only one data point, resulting in an initial transmission close to $\SI{0}{\percent}$, and employed the Upper Confidence Bound (UCB) acquisition function with $\beta = \num{2}$. This algorithm was executed on the machine using an $^{16}$O beam $\num{20}$ times, each time starting from the same conditions. Refer to Fig.~\ref{fig:ATLASnn-prior}, where the blue dashed line represents the median maximum transmission at each step of the constant prior BO optimization, with the surrounding area depicting the $\SI{90}{\percent}$ confidence level. We then trained a neural network model on a dataset comprising $\num{3000}$ experimental data points obtained from a prior experiment involving a $^{14}$N heavy-ion beam to predict transmission as a function of the quadrupole configurations. Doing this, we demonstrate and take advantage of effective transfer learning, where historical data gathered during ATLAS operations using one species, $^{14}$N, is used to inform optimization of accelerator operations for a different but similar heavy-ion species, $^{16}$O. The trained NN model was then incorporated into the BO algorithm as the prior mean and used for optimization with the $^{16}$O beam, employing the same initial conditions as the constant prior BO optimization mentioned earlier. This process was repeated $\num{20}$ times as well. The results, depicted in orange in Fig.~\ref{fig:ATLASnn-prior}, clearly illustrate the superior performance of BO with a neural network prior mean. This approach outperforms traditional BO methods with no prior knowledge, underscoring not only the advantage of incorporating a prior mean model in Gaussian processes but also the potential of transfer learning techniques in optimizing particle accelerators by using models trained under different beams and conditions.

\begin{figure}
\centering
\includegraphics{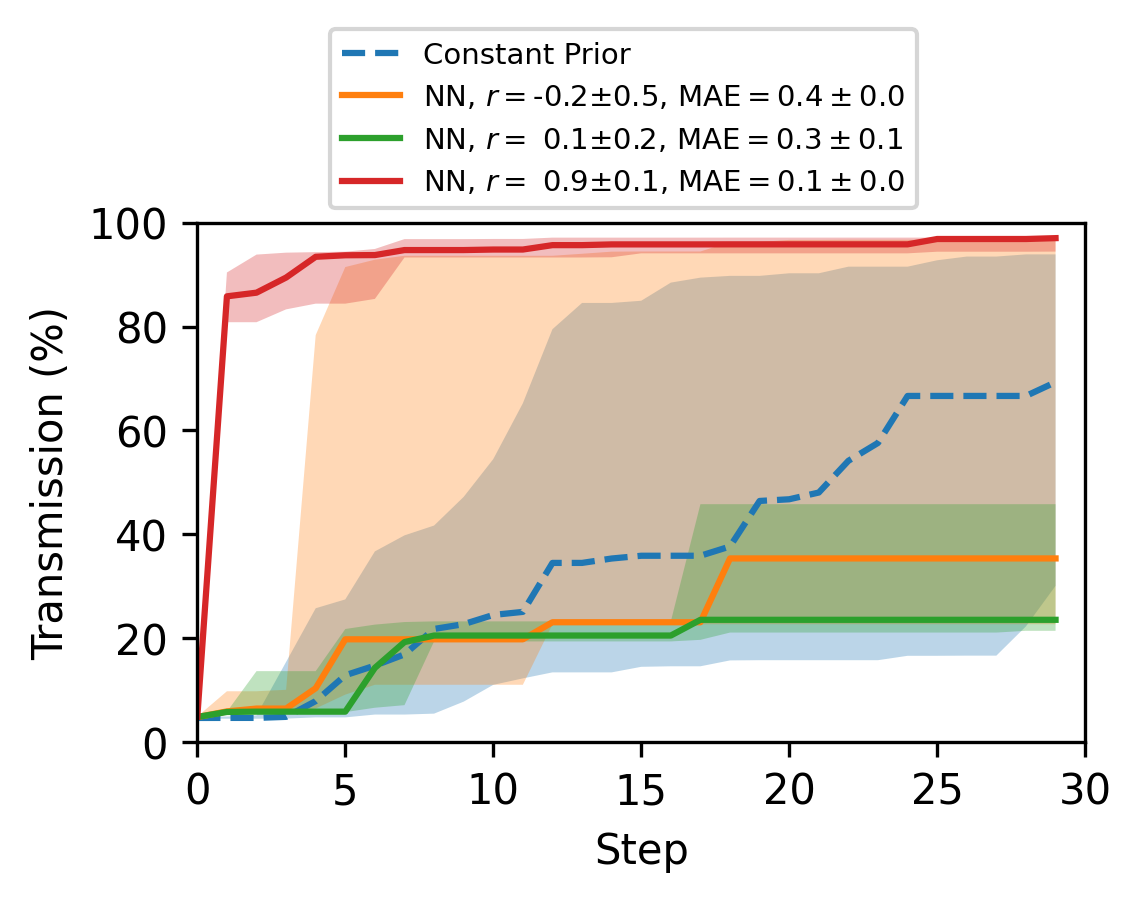}
\caption{Transmission optimization at ATLAS subsection using different prior mean functions. Solid and dashed lines depict the medians and the shaded areas the corresponding $\SI{90}{\percent}$ confidence levels across $\num{10}$ to $\num{20}$ runs.}
\label{fig:ATLAScorr}
\end{figure}

However, as emphasized earlier, the accuracy of the surrogate model holds a pivotal role in determining the effectiveness of the optimization process. This critical factor was empirically validated at ATLAS through an extension of the same BO with GP framework previously mentioned, by employing surrogate models with different accuracies. These surrogate models were trained using the same dataset but with varying duration of training, yielding a spectrum of accuracies measured by different correlations between their predictions and the actual transmission in the machine. Figure~\ref{fig:ATLAScorr} illustrates how the performance of BO augmented by an NN prior mean exhibits marked improvement with the increasing correlation between the model's predictions and the machine's behavior.

We also observe that the use of priors can in some cases lead to widely varying optimization performance, as seen when using the neural network model that has measured correlation coefficients of $r=\num{-0.2}\pm\num{0.5}$. In this case, the measured correlation coefficients of the individual optimization runs have a much larger variance than other cases. One explanation for this could be that for this model, successive optimization runs encounter both regions of parameter space where the prior mean has a strong correlation with the true objective function leading to good performance and regions of parameter space where the model has a poor (or negative) correlation with the true objective, leading to sub-optimal performance. A closer investigation of this large variance in optimization performance is the subject of future study.

\section*{Discussion}
In this work we have experimentally demonstrated how incorporating a fast executing NN surrogate model of the target system can dramatically improve the performance of BO algorithms for two important use-cases: injector optimization at LCLS and switching between ion species at the ATLAS heavy ion linac. We also show in a simulation example that in cases where the prior model is more accurate and shows stronger correlation with ground truth values, the algorithm significantly outperforms standard BO with a constant prior mean function.  These NN based prior mean models can be used to efficiently incorporate information from beam dynamics simulations, historical data, and similar accelerator operating configurations to accelerate online tuning of accelerators, all while quickly adapting to discrepancies between surrogate model predictions and observations on the real machine. However, care must be taken if the prior mean model is poorly correlated or anti-correlated with the true objective function, which can reduce BO performance below the performance with a constant prior. As was observed in the ATLAS experimental demonstration, the use of an expressive prior mean model can also facilitate transfer learning across different beam setups. This can aid rapid switching between beam setups at accelerators that serve a variety of experiments with different requirements (e.g., different species of beams as in the ATLAS case, or different electron beam characteristics requested by photon science users at the LCLS).

Incorporating NN based priors into GP modeling is also a path toward enabling facility-scale BO with hundreds of input parameters. Although these large parameter spaces remain challenging to navigate, including a prior mean model can reduce the complexity by guiding the search to the most relevant areas. In very high-dimensional parameter spaces, the implicit optimization of the acquisition function may eventually also become a non-trivial task.

Neural network surrogate models are able to ingest extremely large sets of historical and simulated data without suffering from an increase in prediction cost, unlike GP models, which scale poorly with the number of training points. By using NN based surrogate models as prior means we can incorporate much more information about the objective function into BO than we would be able to otherwise. Additionally, combining NN surrogates with GP modeling improves the applicability of using NNs in situations where the NN model does not perfectly predict the real objective function. Gaussian process models can quickly adapt predictions to new data, whereas NN surrogates need significant amounts of data and training time to be calibrated to experimental conditions. Combining the two model types leverages the advantages of both modeling strategies and can be used to provide accurate, facility-scale models of objective functions that are updated in real-time.

Future work in this area will focus on studying metrics for training prior models that yield good performance in the context of BO, including disentangling the effects of the two metrics identified in this work to describe model quality, i.e., MAE and correlation.  Additionally, as discussed in the simulation section, it is possible to recover optimization performance on-the-fly, i.e., during optimization, when the prior mean model is a poor predictor of the objective function; future work should more thoroughly explore different heuristics for weighting the use of a NN prior mean function and trainable constant prior functions to improve performance. Furthermore, differentiable beam dynamics models, such as those implemented in Bmad-X~\cite{gonzalez-aguilera_towards_2023} or Cheetah~\cite{kaiser_cheetah_2024,stein_accelerating_2022} should also start to be incorporated into GP models as physics-informed priors. Finally, non-constant prior mean functions can be incorporated into GP models with constraining functions for use in constrained optimization (see~\cite{roussel_bayesian_2024} for details) to reduce the number of constraint violations during optimization.

\section*{Methods}
We begin with a discussion of GP modeling and the method by which we incorporate prior mean functions into the modeling process.
We then discuss how we characterized prior model quality, construction, and software implementation details regarding how NN models were integrated into GP models using the software package Xopt \cite{roussel_xopt_2023}.
\subsection*{Gaussian Process Modeling and Bayesian Optimization}
A Gaussian process~\cite{rasmussen_gaussian_2006} (GP) is a distribution of functions denoted as
\begin{equation}
    f(\mathbf{x})\sim \mathcal{GP}(m(\mathbf{x}), k(\mathbf{x},\mathbf{x^{\prime}})) ~,
    \label{eq:gaussian_process}
\end{equation}
that is fully characterized by its mean function $m(\mathbf{x}) = \mathbb{E}[f(\mathbf{x})]$ and covariance or kernel function $k(\mathbf{x},\mathbf{x^{\prime}}) = \mathbb{E}[(f(\mathbf{x}) - m(\mathbf{x}))(f(\mathbf{x^{\prime}}) - m(\mathbf{x^{\prime}}))]$. It is a generalization of the multivariate Gaussian distribution for which each finite collection of function values $\mathbf{f}$ follows a joint multivariate normal distribution. In applications of BO and to simplify calculations, the prior mean function is commonly specified as a zero mean, i.e., a function that is zero everywhere in the parameter space $\mathcal{X} \subset \mathbb{R}^{d}$
\begin{equation}
    m(\mathbf{x}) = 0 ~\text{for}~\forall \mathbf{x}\in\mathcal{X} ~.
    \label{eq:zero_mean}
\end{equation}
Given a set of $n$ collected data samples $\mathcal{D} = \{X, \mathbf{y}\}$, the posterior distribution evaluated on $n^{*}$ test points can then be expressed as
\begin{equation}
    p(\mathbf{y}_{*} X_{*}, \mathcal{D}) = \mathcal{N}(\bm{\mu}_{*}, \bm{\sigma}_{*}^2)
    \label{eq:posterior_distribution}
\end{equation}
with mean and variance given by
\begin{align}
    \bm{\mu}_{*} &= K(X_{*}, X)[K(X,X) + \sigma_\epsilon^2I]^{-1}\mathbf{y}\label{eq:posterior_mean}\\
    {\bm{\sigma}_{*}}^2 &= K(X_{*},X_{*}) - K(X_{*}, X)[K(X,X) + \sigma_\epsilon^2I]^{-1} K(X_{*},X)^T\label{eq:posterior_variance},
\end{align}
where $K(\cdot\,\cdot)$ denotes the covariance matrix between the respective sets of data and $I$ is the identity matrix. Note that to account for a constant mean function that is not zero, i.e., a function which has a constant, non-zero value everywhere in the parameter space
\begin{equation}
    m(\mathbf{x}) = c ~\text{for}~ \forall \mathbf{x}\in\mathcal{X} ~\text{with}~ c \neq 0,
    \label{eq:constant_mean}
\end{equation}
we can simply redefine the objective function as
\begin{equation}
    f^{\prime}(\mathbf{x}) = f(\mathbf{x}) - c ~,
    \label{eq:constant_prior_mean}
\end{equation}
and still use Eq.~(\ref{eq:posterior_mean}) to compute the posterior mean.

The GP model is then passed to an acquisition function, which uses the model to predict the value of potential future measurements towards finding a global extremum. Points in parameter space that maximize the acquisition function are then observed experimentally and added to the model data set. This process repeats sequentially until a convergence criteria is met. Regardless of the choice of acquisition function, the performance of BO algorithms depends on accurate GP surrogate modeling of the objective function using as few measurements as possible, which we examine in this work.

\subsection*{Non-Constant Prior Means}
If there is sufficient prior knowledge about the objective function to build an \textit{a-priori} model, this initial estimate can serve as a more informative prior mean function for the GP and thereby improve the model's predictive accuracy and BO convergence speed. Incorporating a non-zero prior mean function $m(\mathbf{x})$ into a GP model re-incorporates an extra term ignored in Eq.~\ref{eq:posterior_mean}, producing posterior mean function values
\begin{equation}
    \bm{\mu}_{*} = \mathbf{m}(X_{*}) + K(X_{*}, X)K_y^{-1}(\mathbf{y} - \mathbf{m}(X))
    \label{eq:posterior_mean_for_non_zero_prior}
\end{equation}
with $K_y = K(X,X) + \sigma_\epsilon^2I$. 
For test points that are far away from previous measurements in parameter space ($K(X_{*}, X) \rightarrow 0$), the posterior mean function values $\bm{\mu}_{*}$ are equal to the prior mean values at the test points $\mathbf{m}(X_{*})$. 

\begin{figure}
    \centering
    \includegraphics{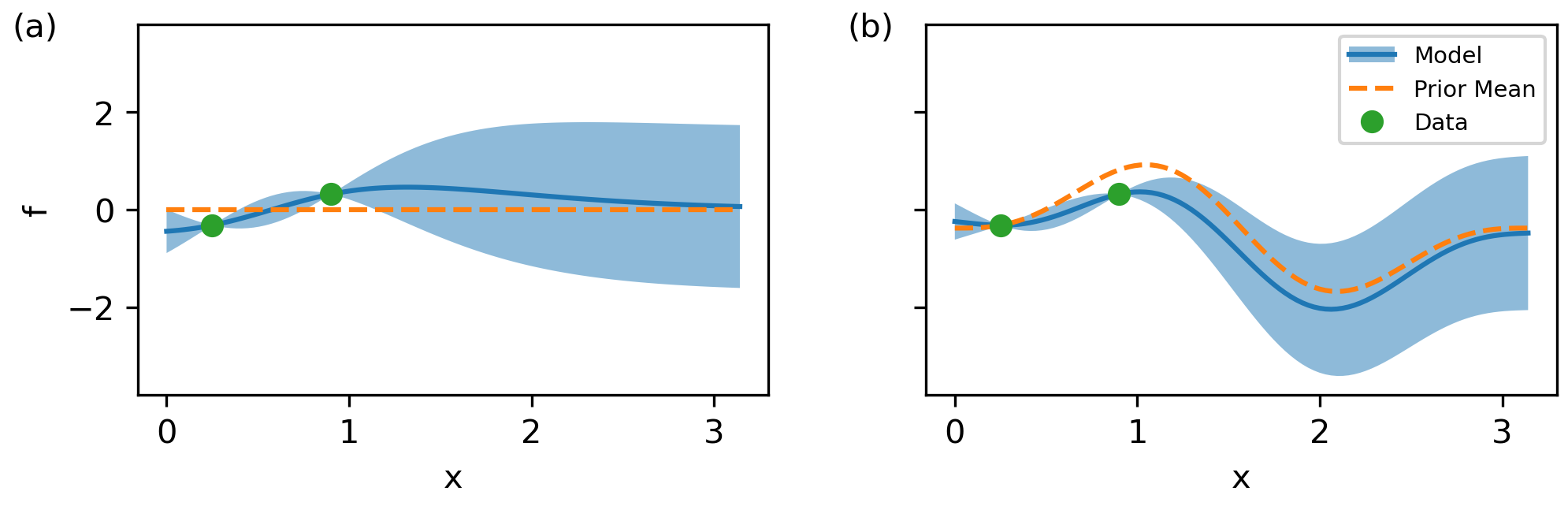}
    \caption{Illustration of non-zero prior mean. In the absence of local data, the mean of the posterior distributions reverts to (a) zero or (b) the non-constant prior mean. The variance remains unchanged.}
    \label{fig:prior_mean_example}
\end{figure}

This effect is illustrated in Fig.~\ref{fig:prior_mean_example}, where the mean of the posterior distribution reverts back to the prior mean as the distance between test points and training data increases. If the prior mean function accurately predicts the objective function, the GP model can make similarly accurate predictions of the objective without any data.  Conversely, if portions of the prior mean incorrectly make predictions, the posterior predictions of the GP model will reflect updated values from training data. In this way, the GP can be interpreted as a model of the difference between the prior mean function $m(\mathbf{x})$ and the true objective function.

\subsection*{Metrics for Characterizing Model Accuracy}
\begin{figure}
    \includegraphics{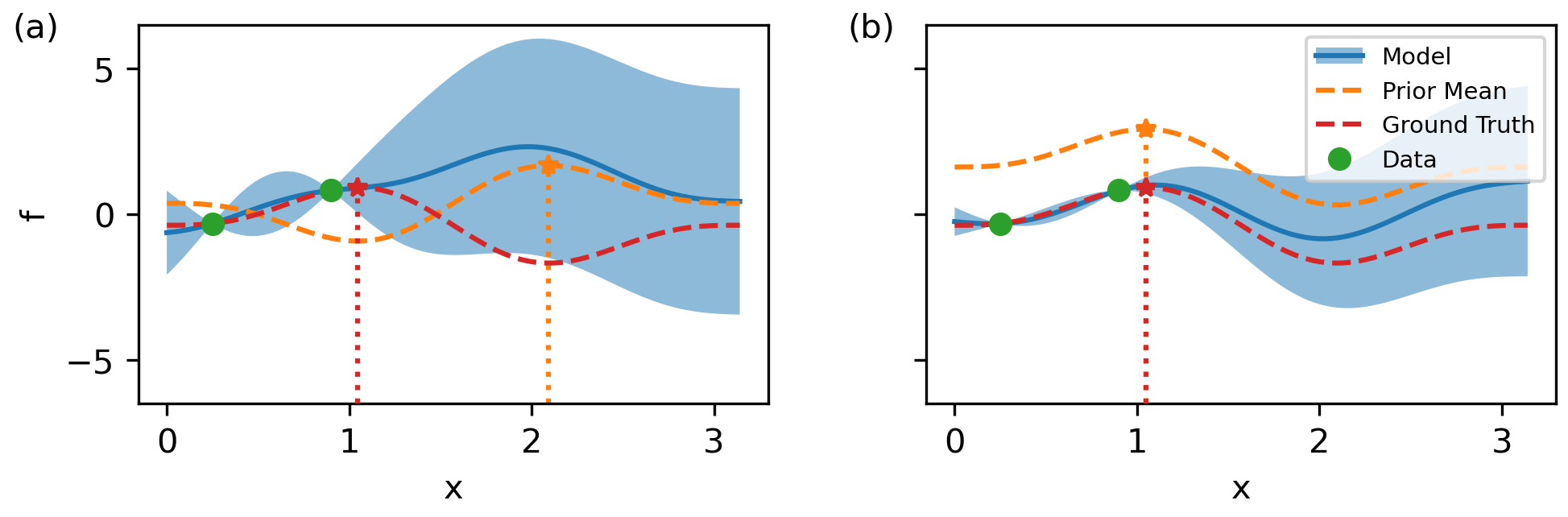}
    \centering
    \caption{Illustration of different metrics for assessing the accuracy of a prior mean model. (a) A prior mean model with a low correlation changes the position of the predicted optimum. (b) In contrast, a model with perfect correlation and a fixed offset retains the position of the optimum, but provides inaccurate value predictions in areas where no data has been observed.}
    \label{fig:prior_metrics_example}
\end{figure}

We use two metrics to quantify the quality of the NN prior model in the context of improving BO performance. While a simple mean squared error (MSE) loss is used in training the NN parameters, this is an insufficient metric to assess the overall model quality for its use as a prior mean function in BO, as illustrated in Fig.~\ref{fig:prior_metrics_example}. We see in Fig.~\ref{fig:prior_metrics_example}(a) that using a prior mean model which has a relatively low MSE, but is poorly correlated with the ground truth objective function, will predict optimal locations that do not correspond to real extrema in the ground truth. However, if the prior mean model strongly correlates with the ground truth function, as shown in Fig.~\ref{fig:prior_metrics_example}(b), predicted optimal locations agree well with the real optimal locations, even if the MSE of the model is larger. In our studies, we use the Pearson correlation coefficient
\begin{equation}
    r = \text{PCC}\,(\mathbf{m}(X),\mathbf{y}) = \frac{\text{cov}(\mathbf{m}(X),\mathbf{y})}{\sigma_{\mathbf{m}(X)}\sigma_{\mathbf{y}}} ~,
    \label{eq:correlation_coefficient}
\end{equation}
where $\text{cov}\,(\cdot\,,\cdot)$ is the covariance and $\sigma_{\mathbf{m}(X)}$ and $\sigma_{\mathbf{y}}$ denote the respective standard deviations. We find this is typically a more accurate metric in assessing model quality and predicting BO performance. This can for example be seen in Fig.~\ref{fig:HighDimensionalBO}, where the model with the highest correlation outperforms other versions, although their absolute error is smaller. Furthermore, as a slightly more intuitive measure of the absolute error than MSE, we augment the correlation coefficient with the mean absolute error (MAE)
\begin{equation}
    \text{MAE}\, (\mathbf{m}(X),\mathbf{y}) = \frac{1}{n} \sum_{i=1}^{n} |y_{i} - m(\mathbf{x}_{i})|
~,
    \label{eq:mean_absolute_error}
\end{equation}
to describe the overall model quality.

\subsection*{Neural Network Models}
\paragraph{LCLS Injector Model}
The LCLS injector model is a fully connected neural network implemented in PyTorch~\cite{ansel_pytorch_2024} with nine hidden layers, including up to $\num{300}$ nodes and Exponential Linear Unit (ELU) activation functions ($\alpha = \num{1.0}$). Furthermore, six dropout modules ($p = \num{0.05}$) are added after the 2nd to 7th layer to help prevent overfitting. The model was trained initially on IMPACT-T simulation results and then calibrated to more closely match the measured data. The simulation data set was generated using Xopt within the parameter ranges listed in Table~\ref{tab:LCLSSimulationParameterRanges}. To study how the model's accuracy impacts its use as a prior mean model, additional low-accuracy variations, listed in Table~\ref{tab:LCLSModelMetrics}, were generated by varying the amount of training time. In order to solve the calibration problem, we froze the weights and biases of the original neural network and added linear nodes to each input. These nodes have free offset and scaling factors, i.e., a weight and a bias, with a linear activation function. These unknowns are then determined by re-training the modified network on measured data in the usual way (stochastic gradient descent and backpropagation through the neural network model). This enables simultaneous determination of calibration factors for multiple inputs and retains interpretability over a non-linear approach. The final model was then deployed as a prior mean model in Xopt within the parameter ranges listed in table~\ref{tab:LCLSExperimentalParameterRanges}.

\begin{table}
\centering
\begin{tabular}{|l|r|r|r|}
\hline
\textrm{Parameter}&
\textrm{Min.}&
\textrm{Max.}&
\textrm{Unit}\\
\hline
Laser Spot Size & $\num{210.2}$ & $\num{500.0}$ & $\si{\micro\meter}$\\
\hline
Laser Pulse Length & $\num{1.818}$ & $\num{7.272}$ & $\si{\pico\second}$\\
\hline
Bunch Charge & $\num{0.250}$ & $\num{0.250}$ & $\si{\pico\coulomb}$\\
\hline
Solenoid 121 & $\num{0.377}$ & $\num{0.498}$ & $\si{\kilo\gauss.\meter}$\\
\hline
Quadrupole 121 & $\num{-0.021}$ & $\num{0.021}$ & $\si{\kilo\gauss}$\\
\hline
Quadrupole 122 & $\num{-0.021}$ & $\num{0.021}$ & $\si{\kilo\gauss}$\\
\hline
Linac 300 Amplitude & $\num{58.0}$ & $\num{58.0}$ & $\si{\mega\volt}$\\
\hline
Linac 300 Phase & $\num{-25.0}$ & $\num{10.0}$ & $\si{\deg}$\\
\hline
Linac 400 Amplitude & $\num{70.0}$ & $\num{70.0}$ & $\si{\mega\volt}$\\
\hline
Linac 400 Phase & $\num{-25.0}$ & $\num{10.0}$ & $\si{\deg}$\\
\hline
Quadrupole 361 & $\num{-4.318}$ & $\num{-1.080}$ & $\si{\kilo\gauss}$\\
\hline
Quadrupole 371 & $\num{1.091}$ & $\num{4.310}$ & $\si{\kilo\gauss}$\\
\hline
Quadrupole 425 & $\num{-7.560}$ & $\num{-1.081}$ & $\si{\kilo\gauss}$\\
\hline
Quadrupole 441 & $\num{-1.078}$ & $\num{7.560}$ & $\si{\kilo\gauss}$\\
\hline
Quadrupole 511 & $\num{-1.079}$ & $\num{7.558}$ & $\si{\kilo\gauss}$\\
\hline
Quadrupole 525 & $\num{-7.560}$ & $\num{-1.080}$ & $\si{\kilo\gauss}$\\
\hline
\end{tabular}
\caption{\label{tab:LCLSSimulationParameterRanges}%
LCLS Simulation Parameter Ranges
}
\end{table}

\begin{table}
\centering
\begin{tabular}{|l|r|r|}
\hline
\textrm{Epochs}&
\textrm{Correlation $r$}&
\textrm{MAE ($\si{\milli\meter}$)}\\
\hline
$\num{1}$ & $\num{-0.13}\pm\num{0.13}$ & $\num{1.50}\pm\num{0.16}$\\
\hline
$\num{2}$ & $\num{0.35}\pm\num{0.08}$ & $\num{1.18}\pm\num{0.11}$\\
\hline
$10$ & $\num{0.69}\pm\num{0.08}$ & $\num{0.63}\pm\num{0.07}$\\
\hline
\end{tabular}
\caption{\label{tab:LCLSModelMetrics}LCLS NN models with different accuracies}
\end{table}

\begin{table}
\centering
\begin{tabular}{|l|r|r|r|}
\hline
\textrm{Name}&
\textrm{Min.}&
\textrm{Max.}&
\textrm{Unit}\\
\hline
Solenoid 121 & $\num{0.479}$ & $\num{0.481}$ & $\si{\kilo\gauss.\meter}$\\
\hline
Quadrupole 121 & $\num{-0.015}$ & $\num{0.019}$ & $\si{\kilo\gauss}$\\
\hline
Quadrupole 122 & $\num{-0.021}$ & $\num{0.006}$ & $\si{\kilo\gauss}$\\
\hline
Quadrupole 361 & $\num{-4.318}$ & $\num{-2.080}$ & $\si{\kilo\gauss}$\\
\hline
Quadrupole 371 & $\num{1.272}$ & $\num{3.846}$ & $\si{\kilo\gauss}$\\
\hline
Quadrupole 425 & $\num{-3.707}$ & $\num{-1.081}$ & $\si{\kilo\gauss}$\\
\hline
Quadrupole 441 & $\num{-1.078}$ & $\num{3.341}$ & $\si{\kilo\gauss}$\\
\hline
Quadrupole 511 & $\num{0.034}$ & $\num{6.944}$ & $\si{\kilo\gauss}$\\
\hline
Quadrupole 525 & $\num{-5.479}$ & $\num{-1.080}$ & $\si{\kilo\gauss}$\\
\hline
\end{tabular}
\caption{\label{tab:LCLSExperimentalParameterRanges}%
LCLS Experimental Parameter Ranges
}
\end{table}

\paragraph{ATLAS Model}
The neural network model employed to predict transmission in the ATLAS accelerator, which takes five quadrupole currents as input, consists of two hidden layers, each with $\num{20}$ units, and employs the hyperbolic tangent as the activation function. The output layer utilizes a sigmoid function, constraining the transmission values to the $[0, 1]$ range. The model was implemented using PyTorch, with the MSE loss function employed for training. For optimization, the Adam optimizer was used with a learning rate of $\alpha = \num{0.01}$. Additionally, a learning rate scheduler was applied to adaptively adjust the learning rate during training based on the performance improvement. The primary objective of implementing this model was to demonstrate that a simple approach for the prior mean could yield benefits in Bayesian optimization, as detailed in this paper. The model was trained using a dataset comprising $\num{3000}$ experimental data points obtained from a prior experiment involving a $^{14}$N ion beam. Subsequently, the trained model was deployed and evaluated on a different ion species, $^{16}$O, for another experiment. This process showcased the model's capability to effectively transfer knowledge between different ion beams, highlighting its versatility and generalizability. Different models were trained to different levels of fidelity by conducting the training up to various epochs. The respective model fidelity is reflected in the correlation coefficient and MAE listed in Table~\ref{tab:ATLASModelMetrics}.

\begin{table}
\centering
\begin{tabular}{|l|r|r|}
\hline
\textrm{Epochs}&
\textrm{Correlation $r$}&
\textrm{MAE ($\si{\milli\meter}$)}\\
\hline
$\num{1}$ & $\num{0.09}\pm\num{0.17}$ & $\num{0.25}\pm\num{0.08}$\\
\hline
$\num{10}$ & $\num{-0.21}\pm\num{0.48}$ & $\num{0.39}\pm\num{0.04}$\\
\hline
$\num{100}$ & $\num{0.88}\pm\num{0.11}$ & $\num{0.05}\pm\num{0.00}$\\
\hline
\end{tabular}
\caption{\label{tab:ATLASModelMetrics}%
ATLAS NN models with different accuracies
}
\end{table}

\subsection*{Practical Implementation of Neural Network Priors for Online Control}
\begin{figure}
\centering
\includegraphics[width=\linewidth]{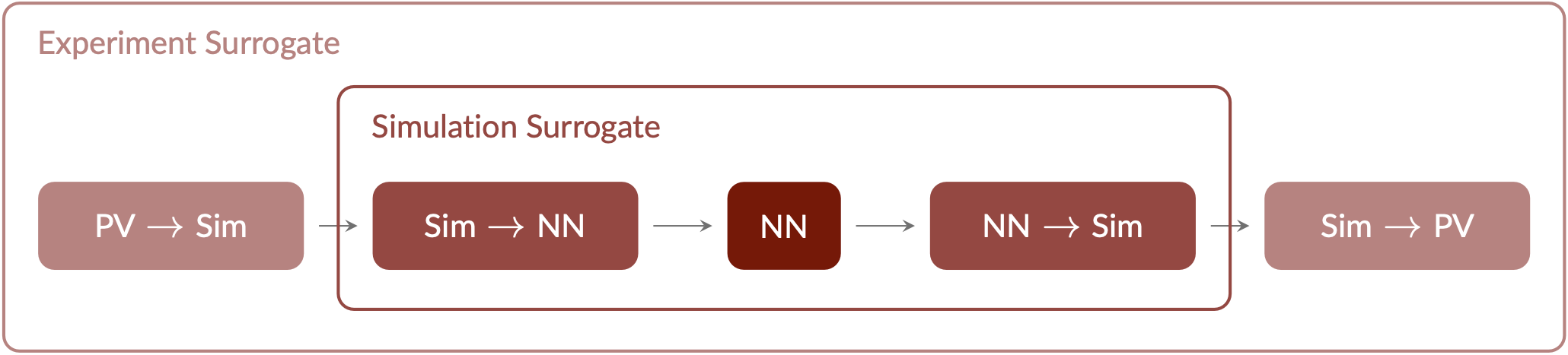}
\caption
{Diagram showing transformation layers implemented in LUME-Model to transform real accelerator parameters to NN inputs.}
\label{fig:transformation_layers}
\end{figure}

While the mathematics of incorporating neural networks into GP models is straightforward, the practical implementation in the context of online accelerator control can be challenging. For example, as discussed above, the LCLS NN surrogate model was trained using simulation data and there is a significant mismatch between the names and units of the simulation parameters and those of the control parameters at the real machine. Translating the inputs and outputs of this model required a number of transformation layers, shown in Fig.~\ref{fig:transformation_layers}, to map control parameters adjusted during optimization to NN input parameters. To address this, we implemented a wrapper class in the LUME-Model package~\cite{mayes_lightsource_2021} which stores the input and output ordering and the required transformations to make the model applicable to the real machine.

Additionally, specialized modules are needed to enable usage of these surrogate models in the Bayesian optimization package BoTorch~\cite{balandat_botorch_2020}. To improve modeling robustness, BoTorch applies data normalization and standardization transformations to input and output data respectively. As a result, incorporating surrogate-based prior mean functions into the GP models created by BoTorch requires specialized modules to un-transform the input data before passing it to the surrogate model and transforming the output data. These modules are built into the Xopt~\cite{roussel_xopt_2023} package for convenient use.

Finally, as the models used in this work are instances of a PyTorch module (\verb|torch.nn.Module|) they register all trainable sub-modules by default. One benefit of this is that custom parameters can easily be optimized as part of the GP model fit by simply activating their gradients. However, as the full GP model is set in training mode before the fit, this will also affect all sub-modules, including the NN prior model. If the NN architecture includes layers like dropout or other modules which change behavior based on whether the model is in training or evaluation mode, this can also affect the prior mean predictions. To address this potential pitfall, modules used as prior mean functions in Xopt are, by default, always queried in evaluation mode.

By creating these open-source software solutions to facilitate easy incorporation of prior models into Bayesian optimization in a reliable way, we are helping to make this technique more broadly accessible to the accelerator community. This reduces development overhead for new facilities and makes the algorithm more ready for use out-of-the-box by accelerator facilities, especially for the growing number that already have available neural network models of their systems.

\section*{Data Availability}
The data sets generated during the current study are available from the corresponding author on reasonable request.


\section*{Acknowledgments}
This work was funded by the U.S. Department of Energy, Office of Science, Office of Basic Energy Sciences under Contract No. DE-AC02-76SF00515. It was also supported by the U.S. Department of Energy, Office of Nuclear Physics under Contract No. DE-AC02-06CH11357. The presented research used the ATLAS facility, which is a DOE Office of Science User Facility.

\section*{Author contributions statement}
Conceptualization, T.B., J.L.M., B.M., D.R., R.R., and A.L.E.; Data curation, T.B., J.L.M., C.X., K.R.L.B., and Z.Z.; Formal analysis, T.B.\ and J.L.M.; Funding acquisition, D.R., A.L.E., and B.M.; Investigation, T.B.\ and J.L.M.; Methodology, R.R., T.B., J.L.M., and A.L.E.; Software, R.R., C.X., K.R.L.B., J.M., T.B., and J.L.M.; Experiment, T.B., J.L.M., R.R., and A.L.E.; Supervision, D.R., A.L.E., R.R., and B.M.; Validation, T.B.\ and J.L.M.; Visualization, T.B.\ and J.L.M.; Writing---original draft, T.B., J.L.M., R.R., B.M., and A.L.E.; Writing---review and editing, T.B., J.L.M., R.R., D.R., B.M., and A.L.E. All authors have read and agreed to the published version of the manuscript.

\section*{Additional information}
\subsection*{Competing Interests}
The author(s) declare no competing interests.

\end{document}